\pdfoutput=1
\documentclass[%
 reprint,
 amsmath,amssymb,
 aps
]{revtex4-2}
\usepackage{lipsum}
\usepackage{graphicx} % Required for inserting images
\usepackage{xcolor}
\usepackage{amsmath}
\usepackage[colorlinks=true, allcolors=blue]{hyperref}
\usepackage[normalem]{ulem}

\makeatletter
\renewcommand\@fnsymbol[1]{\ensuremath{\ifcase#1\or \dagger\or \ddagger\or
  \mathsection\or \mathparagraph\or *\or **\else\@ctrerr\fi}}
\makeatother

\date{March 2026}

\begin{document}
\title{Avalanche Size and Interevent Time Statistics in Amorphous Systems}
\author{Thomas Muhren$^1$}
\author{Roberto Benzi$^{2,3}$}
\email[Contact Author: ]{roberto.benzi@gmail.com}
\author{Mauro Sbragaglia$^3$}
\author{Jeannot Trampert$^4$}
\author{Federico Toschi$^1$}
\affiliation{$^{1}$Department of Physics, Eindhoven University of Technology, \\PO Box 513, 5600  MB, Eindhoven, The Netherlands.}
\affiliation{$^{2}$ Sino-Europe Complex Science Center, School of Mathematics, North University of China, Shanxi 030051, Taiyuan, China}
\affiliation{$^{3}$ Department of Physics, University of Tor Vergata and INFN, Via della Ricerca Scientifica 1, 00133 Rome, Italy.}
\affiliation{$^{4}$ Department of Earth Sciences, Utrecht University, \\ Princetonlaan 8a, 3584 CB Utrecht, The Netherlands.}

\begin{abstract}
 The intermittent, stick-slip flow of amorphous systems is characterized by avalanches whose statistical properties display robust power-laws. In this paper, we study avalanche statistics in a field theoretical model of a soft-glassy system under shear. The order parameter in our model is the fluidity, which is related to the local rate of plastic events in a soft-glass. While avalanche sizes $S$ have been extensively characterized, the statistics of interevent times $t_i$ have remained largely unexplored. Here, we find that both $S$ and $t_i$ are power-law distributed and that their scaling exponents satisfy a precise relation. The model enables the analytical derivation of the scaling exponents and the reproduction of the behavior observed across a wide class of amorphous systems.
\end{abstract}

\maketitle
\begin{figure}[h!]
    \centering
    \includegraphics[width = 0.99\linewidth]{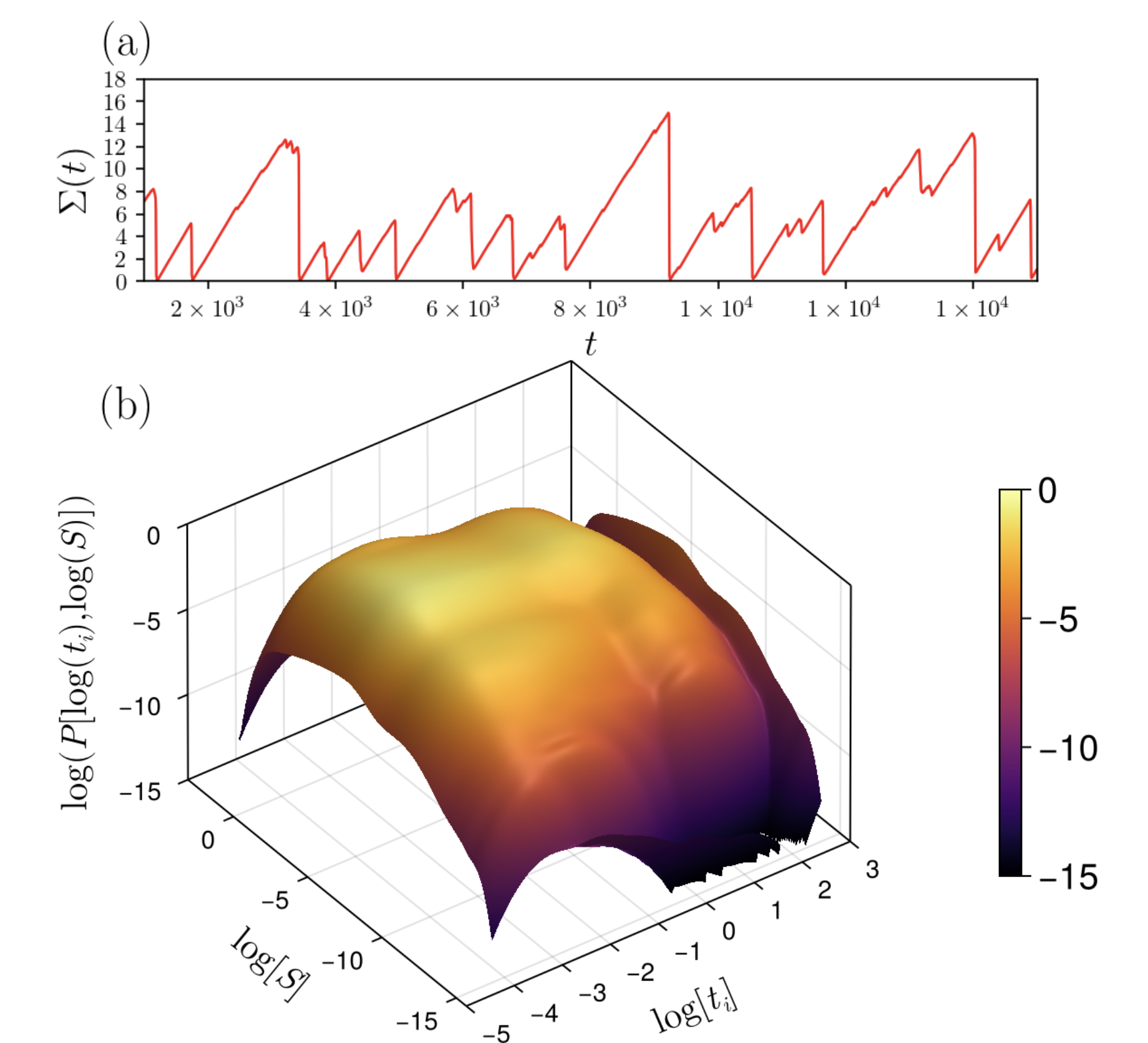}
    \caption{ Panel (a): Stress signal $\Sigma(t)$ as a function of time $t$ for the space-independent model given by Eqs.~\eqref{eq:maxwell_model} and~\eqref{eq:0d_fluidity}. Panel (b): joint probability density of interevent times and sizes. The Pearson correlation coefficient, $r=0.003$ indicates the two are essentially uncorrelated.
    Numerical results are obtained with: $R=0.016$, $\epsilon_1=0.1, \tau=1$ and $\epsilon_0 = 10^{-5}$(a) and $\epsilon_0 = 10^{-7}$(b). 
    }
    \label{fig:stress_timeseries}
\end{figure}

The dynamics of avalanches can be observed in many different physical systems. Examples include amorphous solids, soft-glasses, sand piles, solar flares, earthquakes and the random-field Ising model~\cite{Sethna_Dahmen_Myers_2001,  Uhl_Pathak_Schorlemmer_Liu_Swindeman_Brinkman_LeBlanc_Tsekenis_Friedman_Behringer_etal._2015, Dahmen_Ben-Zion_Uhl_2009, Denisov_Lorincz_Wright_Hufnagel_Nawano_Gu_Uhl_Dahmen_Schall_2017, Boffetta_Carbone_Giuliani_Veltri_Vulpiani_1999}. 
The peculiar signature common to all of these systems is the power-law behavior of the avalanche size $S$ distribution $P[S] \sim S^{-\alpha}$, where usually $S$ is defined using some threshold $M_t$ depending on the physical property of the system~\cite{Davidsen_Kwiatek_2013,Lherminier_Planet_Vehel_Simon_Vanel_Maloy_Ramos_2019}. The exponent $\alpha$ depends on the dimensionality of the system and, obviously, on the physical mechanism underlying the dynamics~\cite{Lin_Lerner_Rosso_Wyart_2014}. Many theoretical studies have been conducted to compute this exponent, and many investigations have identified power-law behavior as criticality, following the pioneering work on self-organized criticality by Bak {\it et al.}~\cite{Bak_Tang_Wiesenfeld_1988}. Another interesting quantity is the avalanche duration time, $t_E$, which is always correlated with avalanche size. Few approaches have been developed to study the interevent time $t_i$ between two successive avalanches. Within the framework of self-organized criticality, $t_i$ is independent of the avalanche size, $S$, and is Poisson distributed~\cite{Bak_Tang_Wiesenfeld_1988}. Conversely, experimental and numerical results have shown that $t_i$ can exhibit a power-law distribution~\cite{Davidsen_Stanchits_Dresen_2007, Davidsen_Kwiatek_2013,Lherminier_Planet_Vehel_Simon_Vanel_Maloy_Ramos_2019}. A notable finding across many studies is that the power-law behavior of the  probability distribution of $t_i$ remains unchanged by coarse-graining in avalanche size~\cite{Corral_2004,Kumar_Korkolis_Benzi_Denisov_Niemeijer_Schall_Toschi_Trampert_2020}. This means that by considering only events where $S > \lambda S_0$, we can compute the interevent time distribution,  denoted by $P[t_i|\lambda S_0]$, 
which exhibits power-law behavior with the same exponent as for $\lambda=1$. This differs from the statistics of $t_{i}$ as the threshold $M_{t}$ changes, where the functional form of the probability distribution $P[t_i\vert{}M_t]$  changes under the transformation $M_t \rightarrow \lambda M_t$, becoming exponential in the limit \(\lambda \gg 1\)~\cite{Janicevic_Laurson_Maloy_Santucci_Alava_2016}. In a recent paper~\cite{Benzi_Castaldi_Toschi_Trampert_2022} it has been argued that, if the scaling $P[t_i|\lambda S_0] \sim t_i^{-\gamma}$ is invariant for $\lambda>1$, then
the scaling exponent  ${\gamma}$ should be linked to the scaling exponent $\alpha$ of avalanche distribution,  in excellent agreement with published results~\cite{Corral_2004,Kumar_Korkolis_Benzi_Denisov_Niemeijer_Schall_Toschi_Trampert_2020}. Yet, no theoretical approach has been developed to compute the exponent $\gamma$. The purpose of this letter is to fill this gap. 
We introduce a simple model for avalanche dynamics in soft-glassy systems where both exponents $\alpha$ and $\gamma$ can be computed theoretically. The model shows invariance of the scaling properties of $P[t_i|\lambda S_0]$ by increasing $\lambda$ and, for a suitable choice of parameters, it reproduces the observed scaling properties of many amorphous systems. 
\\
To model a soft glass we will use the framework introduced by Bocquet {\it et al.}~\cite{Bocquet_Colin_Ajdari_2009} and further developed in Benzi {\it et al.}~\cite{Benzi_Sbragaglia_Bernaschi_Succi_Toschi_2016}. 
We consider the case of a Couette geometry between two walls at $y=0$ and $y=L$, where we apply a velocity $U=0$ at $y=0$ and $U = \dot \Gamma L$ at $y=L$, where $\dot \Gamma$ is the shear rate.
The order parameter of the model is the fluidity $f$ linked to the local rate of plastic events. In the following $f$ is supposed to depend on $y$ and on time $t$. Since $f>0$,  following~\cite{Benzi_Sbragaglia_Bernaschi_Succi_Toschi_2016} it is useful to introduce the field $\phi$ such that $f=\phi^2$. Then the equations of motion are
\begin{eqnarray}
 \label{eq:maxwell_model}
    \frac{d\Sigma}{d t} &=& \frac{1}{\tau} \left( {\dot \Gamma} - \langle\phi^2\rangle  \Sigma   \right), \\
\label{eq:fluidity_eq}    
 \frac{\partial \phi}{\partial t} &=& 2\xi^2 \phi^2 \Delta \phi
+2\xi^2 \phi (\nabla \phi)^2 +m\phi^3 - \phi^3 |\phi|\\
\nonumber
    &+&\sqrt{\epsilon_0}\eta_1(y,t) + \sqrt{\epsilon_1 \langle\phi^2\rangle} \eta_2 (t). 
\end{eqnarray}
Eq.~\eqref{eq:maxwell_model} corresponds to the Maxwell equation which describes a viscoelastic material with a stress relaxation part whose characteristic inverse time comes from another dedicated Eq.~\eqref{eq:fluidity_eq}. In the latter equation,
 $\xi$ is the cooperative length related to the non-locality of the flow,  $\eta_1(y,t)$ is a delta correlated Gaussian noise in space and time, $\eta_2(t)$ is a Gaussian noise delta correlated in time and $\langle .. \rangle$ stands for average in $y$. We also define the dimensionless stress as $\Sigma \equiv \sigma/\sigma_y$,  where $\sigma$ is the space averaged stress, and $\sigma_y$ is the yield stress of the system. In Eq.~\eqref{eq:fluidity_eq} the quantity $m$ is  defined as $m^2(\Sigma)= \frac{(\Sigma - 1)^{1/n}}{\Sigma } \Theta(\Sigma - 1)$ , with $\Theta$ being the Heaviside function and $n$ the Herschel-Bulkley exponent. In this continuum framework, the stochastic term proportional to $\epsilon_{0}$ accounts for microstructural fluctuations. Crucially, long-range correlations are enhanced by the multiplicative noise through the $\epsilon _{1}$-dependent term, triggering abrupt, avalanche-like fluidization. More extensive details on the model derivation and properties are recalled in Sec. \ref*{sec:model_equations} of the Supplementary Material (SM). For $\epsilon_1=\epsilon_0=0$, the model defined in Eqs.~\eqref{eq:maxwell_model} and~\eqref{eq:fluidity_eq} has successfully been used to quantitatively explain the scaling of the fluidization time in shear and stress start up experiments~\cite{Benzi_Divoux_Barentin_Manneville_Sbragaglia_Toschi_2019} and to predict the scaling properties of stress overshoot as a function of $\dot \Gamma$~\cite{PhysRevLett.127.148003}. For $\epsilon_0$ and $\epsilon_1$ different from zero, the model is also able to describe known results for avalanches and brittle yielding transitions~\cite{Benzi_Divoux_Barentin_Manneville_Sbragaglia_Toschi_2021}. Following~\cite{Benzi_Divoux_Barentin_Manneville_Sbragaglia_Toschi_2021}, we
assume $\partial_y f=0$ for $y=0,L$ and we discuss in  Sec.~\ref*{sec:model_equations} of SM the effect of different choices of boundary conditions. Finally,  in all our simulations we assume $\dot \Gamma =0.01$, $n=1/2$ and $L=1$.
\\
For our purposes we can greatly 
simplify Eq.~\eqref{eq:fluidity_eq} in two steps. First, 
following Ref.~\cite{Benzi_Sbragaglia_Bernaschi_Succi_Toschi_2016} it is possible to show that Eq.~\eqref{eq:fluidity_eq} can be reduced to 
the following nonlinear diffusion equation (see Sec.~\ref*{sec:self-consistent-approx} of SM for details)
\begin{equation}\label{eq:r_fluidity}
    \frac{\partial \phi}{\partial t} =  D \partial_y
    ^2 \phi-R \phi + m(\Sigma)\phi^3 - \phi^3 |\phi| + \sqrt{\epsilon_0 + \epsilon_1 \langle \phi^2\rangle} \eta(t),
\end{equation}
where $R=\langle 2\xi^2 (\partial_y \phi)^2\rangle$ and $D=\langle 2 \xi^2 \phi^2 \rangle$ are renormalized coefficients induced by fluctuations that 
depend on $\epsilon_0$ and $\epsilon_1$. To first order in $\epsilon_{1}$, the correlation scale $l_c \equiv \sqrt{D/R}$ can be expressed as (see Sec.~\ref*{sec:self-consistent-approx} of SM): \begin{equation}\label{eq:lc}
l_c = \frac{l_0}{\sqrt{1-1/(2\delta)}}
\end{equation}
where the subscript $0$ labels quantities evaluated at $\epsilon_1 = 0$, $\delta\equiv R_0/\epsilon_1$, and $l_0=\sqrt{D_0/R_0} \sim k_M^{-1}$ The parameter $k_{M}^{-1}$ serves as a regularization lengthscale corresponding to the minimum mesoscopic lengthscale of the system (see Sec.~\ref*{sec:self-consistent-approx}. of SM). Eqs.~\eqref{eq:maxwell_model}-\eqref{eq:fluidity_eq} are invariant under the scale transformation: 
$ m \rightarrow \lambda m$, $\phi \rightarrow \lambda \phi$, $\Sigma \rightarrow \lambda^2 \Sigma$, $t \rightarrow \lambda^{-3} t$, $\tau \rightarrow \lambda^{-1} \tau$, $\epsilon_1 \rightarrow \lambda ^3 \epsilon_1$ (see Sec.~\ref*{sec:model_equations} of SM).
The same is true for Eq.~\eqref{eq:r_fluidity} with the scaling $R \rightarrow\lambda^3 R$. Thus, the ratio $R/\epsilon_1$ can be considered as the ratio of two time scales: $R^{-1}$ which
corresponds to the relaxation time near the equilibrium state $\phi=0$ and $\epsilon_1^{-1}$ which is the time required for the system to build the correlation $l_c$. For
$\delta \rightarrow 1/2$ the correlation scale $l_c$ diverges, implying that the system can be described by a space-independent fluidity, such that we can neglect the term $D\partial_y^2 \phi $ in Eq.~\eqref{eq:r_fluidity}. These simplifications result in  the following equation for $\phi=\phi(t)$:
\begin{equation}\label{eq:0d_fluidity}
    \frac{d \phi}{d t} =  -R \phi + m(\Sigma)\phi^3 - \phi^3 |\phi| + \sqrt{\epsilon_0 + \epsilon_1 \phi^2} \eta(t).
\end{equation}
For simplicity, we use $R$ as a tunable parameter, such that for Eq. \ref{eq:0d_fluidity} we redefine $\delta = \frac{R}{\epsilon_1}$. 
From Eqs.~\eqref{eq:r_fluidity} and~\eqref{eq:0d_fluidity}, we observe that the quantity $\phi$ has a stable fixed point at $\phi=0$. A straightforward computation shows that there exists a critical value of $m=m_c \sim R^{1/3}$ such that for $m \ge m_c $ a new stable fixed point appears near $\phi=m$ together with an unstable fixed point close to $\phi \sim \sqrt{R/m}$. Then, it is relatively intuitive to understand the dynamics of the model: starting from $\phi=0$, the value of stress $\Sigma$ increases linearly in time and $m$ can become larger than $m_c$. In this case the noise can eventually induce large fluctuations  in $|\phi|$ up to order $m$. When this happens the second term of Eq.~\eqref{eq:maxwell_model} becomes
much larger than the first one (i.e. $\dot \Gamma$)  producing a sudden drop in stress of the system corresponding to an {\it avalanche}  and causing $\phi$ to return to the fixed point $\phi = 0$. If $\dot \Gamma$ is large enough, then Eq.~\eqref{eq:maxwell_model} exhibits a stationary solution for  $m \ge m_c$ and avalanche events are suppressed. Eq.~\eqref{eq:0d_fluidity} is our simplified model  whose statistical properties we investigate in the following. In Sec.~\ref*{sec:space_dependent_statistics} of SM we report the numerical results obtained using Eq.~\eqref{eq:fluidity_eq} which agree extremely well with the dynamics obtained with the simplified Eq.~\eqref{eq:0d_fluidity}. Fig.~\ref{fig:stress_timeseries}(a)  shows $\Sigma(t)$ as  a function of $t$ obtained by integrating Eq.~\eqref{eq:maxwell_model} together with Eq.~\eqref{eq:0d_fluidity}. 
It can be seen that the stress dynamics is characterized by the formation of avalanches of different sizes and a clear signature of different timescales. In particular, there are relatively long time scales characterizing the stress growth whereas very short timescales are observed during   stress drops or avalanches. During the interevent time between two avalanches the stress rises almost linearly as expected from Eq.~\eqref{eq:maxwell_model}. Notice that the avalanche sizes and the interevent times are not correlated as shown in Fig.~\ref{fig:stress_timeseries}(b). This is quantitatively supported by the Pearson correlation coefficient, which is $r = 0.003$ (see Sec.~\ref*{sec:correlation} of SM for details).

%%%%%%%%%%%%%%%%%%%%%%%%%%
%%%%%%%%%%%%%%%%%%%%%%%%%%%%%
From Fig.~\ref{fig:stress_timeseries}(a) we  define an avalanche when $\frac{d\Sigma}{dt} < 0$, and following~\cite{Kumar_Korkolis_Benzi_Denisov_Niemeijer_Schall_Toschi_Trampert_2020} we define the size of the corresponding avalanche as 
\begin{equation}\label{eq:avalanche_size}
    S = -\int_{t_E} \Sigma \left(   \frac{d\Sigma}{dt}\right)\,dt,
\end{equation}
where $t_E$ is the time duration of the avalanche. This implies that the avalanche size is proportional to the energy released during an event. The magnitude of an event is conventionally defined as being proportional to the logarithm of that energy, i.e. proportional to $\log S$. Using the scale invariant transformation previously described, we find that $S \rightarrow\lambda^4S$ and hence the scaling $S \sim \phi^4$.  By solving the stationary Fokker-Planck equation related to Eq.~\eqref{eq:0d_fluidity} (see Sec. \ref*{sec:probability_distributions} of SM), we get the probability distribution of the fluidity $f=\phi^2$:
\begin{equation} \label{eq:pdf_fluidity}
    P[f] \sim \frac{C}{f^{1/2} \left(1 + A f\right)^{(1+\delta)}}\exp(-N(f))
\end{equation}
where $A = \frac{\epsilon_1}{\epsilon_0} \gg 1$, $C$ is a normalizing factor  and $N(f)$ stands for nonlinear terms in $f$. Then, using $S\sim f^2$,  we obtain the probability distribution of the avalanche size:
\begin{equation} \label{eq:pdf_sizes}
    P[S] \sim  \frac{1}{S^{5/4 + \delta/2}},
\end{equation}
holding in the limit of large avalanches and neglecting the exponential cut-off due to finite size effects (see Sec.~\ref*{sec:probability_distributions} of SM for further details).

%%%%%%%%%%%%%%%%%%%%%%%%%%%%%%%%%%%%%%%%%%%%%%%%%%%
\begin{figure}[h!]
    \centering
    \begin{minipage}{0.99\linewidth}
        \centering
        \includegraphics[width=\linewidth]{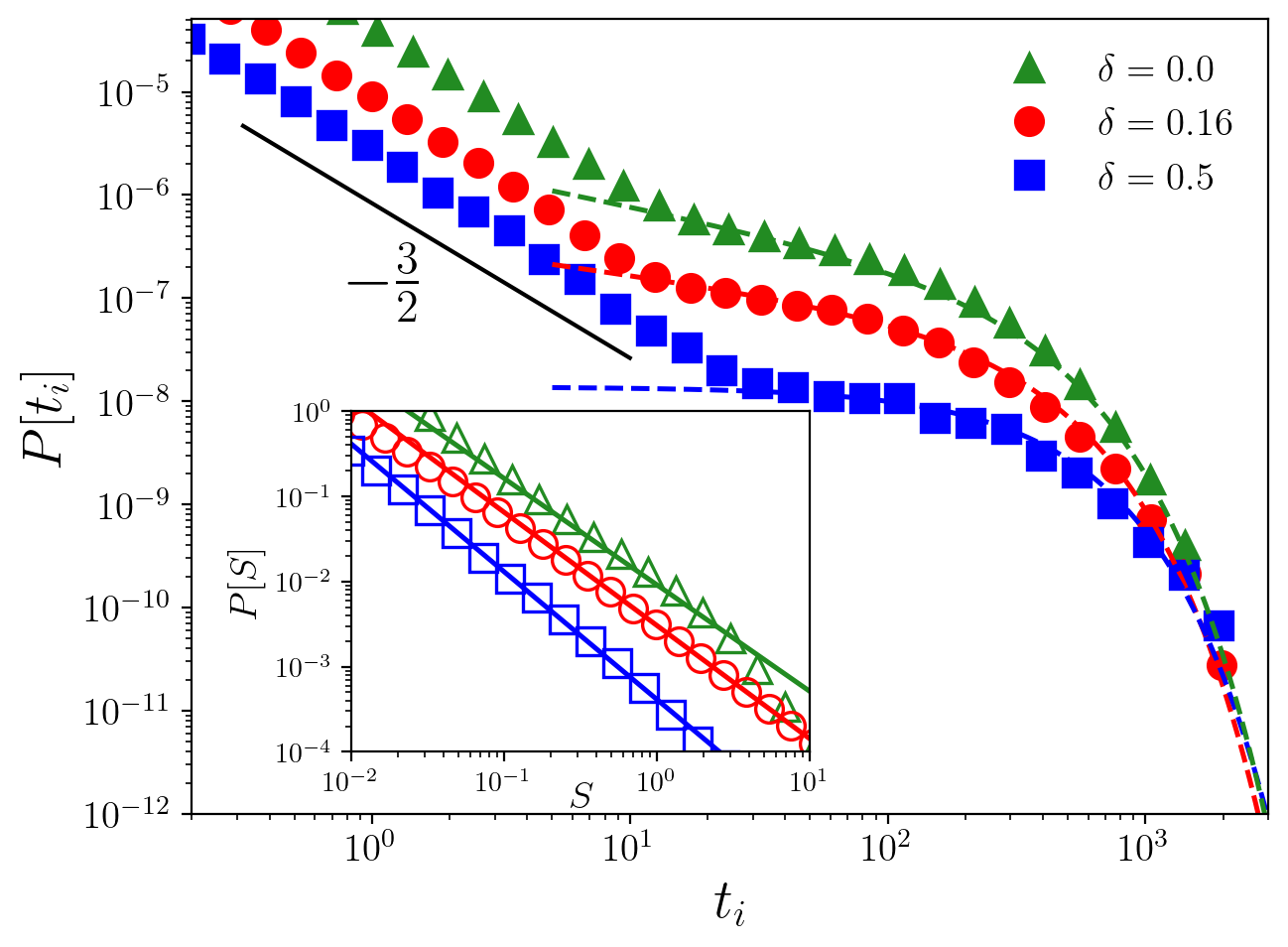}
    \end{minipage}
    \caption{
    Main Panel: Probability density $P[t_i]$ of the interevent time between consecutive avalanches for the fluidity model, Eqs.~\eqref{eq:maxwell_model} and~\eqref{eq:0d_fluidity}. The analytical predictions for $P[t_i]$, given by Eqs.~\eqref{eq:A} and~\eqref{eq:pdf_ti}, are also reported.  Inset: we report the probability density $P[S]$ for the avalanche size together with the analytical prediction given by Eq.~\eqref{eq:pdf_sizes}. The same parameters as in Fig.~\ref{fig:stress_timeseries}(b) have been used.}
    \label{fig:pdf_ti_sizes}
\end{figure}
\begin{figure}
    \centering
    \includegraphics[width=0.97\linewidth]{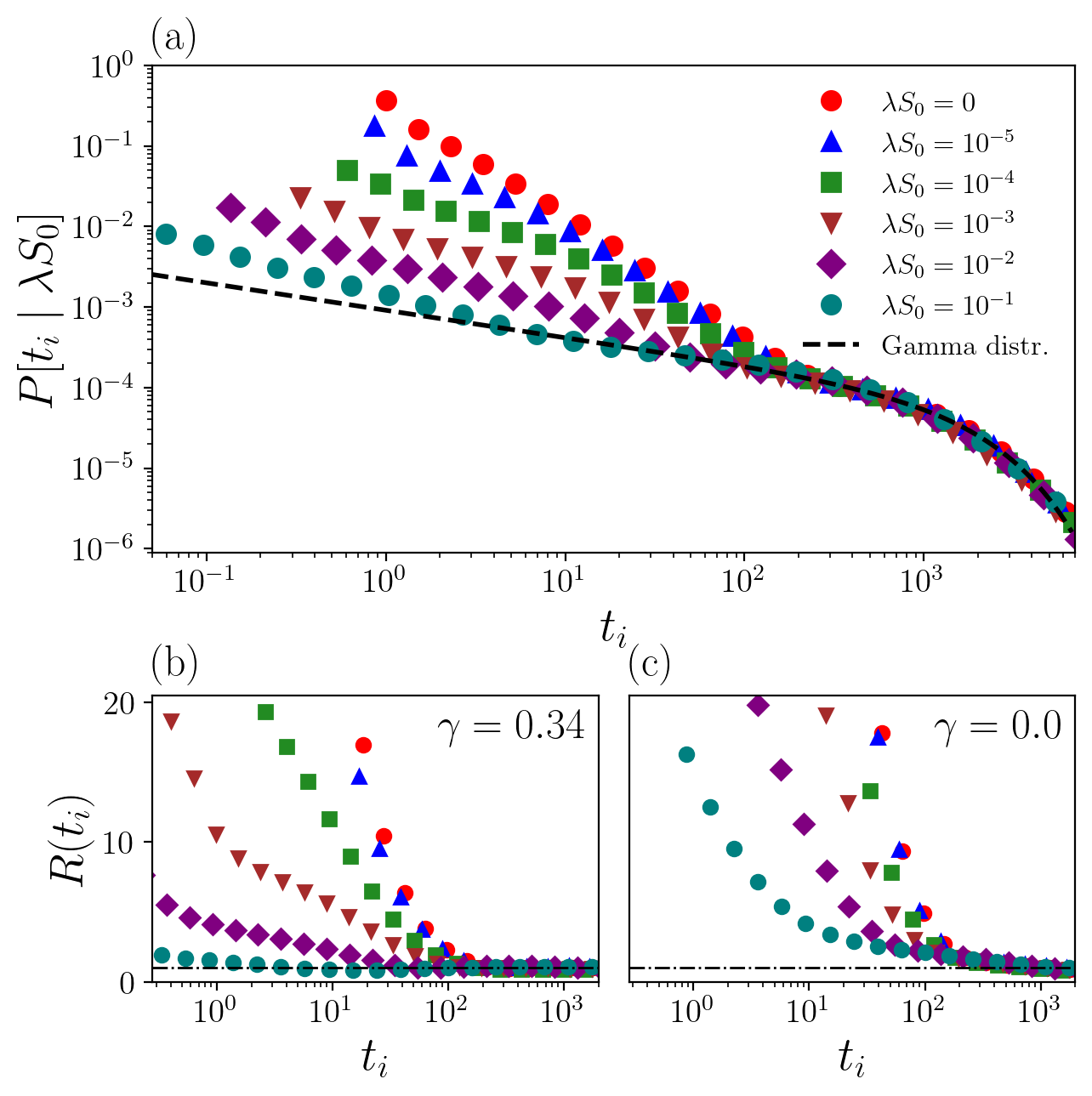}
    \caption{
    Panel (a): probability distribution $P[t_i | \lambda S_0]$ for different values of $\lambda S_0$. 
    The probability distributions collapse on the same master curve, which is the Gamma distribution with $\gamma = 0.34$.  We report the ratio $R(t_i)$ with the Gamma distribution for $\gamma = 0.34$ (b) and $\gamma = 0.0$ (c). Results obtained with the same parameters as in Fig.~\ref{fig:stress_timeseries}(b) with $\delta = 0.16$.}
    \label{fig:0d_coarsening}
\end{figure}

%%%%%%%%%%%%%%%%%%%
This prediction matches the numerical results for $\delta=0,0.16,0.5$ as shown in the inset of Fig.~\ref{fig:pdf_ti_sizes}. In the latter case we obtain the mean-field prediction with $P[S] \sim S^{-1.5}$~\cite{Dahmen_Ben-Zion_Uhl_2009,Denisov_Lorincz_Wright_Hufnagel_Nawano_Gu_Uhl_Dahmen_Schall_2017,Lin_Lerner_Rosso_Wyart_2014}. In Sec.~\ref*{sec:critical_signal} of SM we checked the validity of Eq.~\eqref{eq:pdf_sizes} for different $\delta$ using Eq.~\eqref{eq:fluidity_eq} for the numerical simulations.
\\
Next we turn our attention to the probability distribution of the interevent time $t_i$. Looking at Fig.~\ref{fig:stress_timeseries} and
using Eq.~\eqref{eq:maxwell_model}, we can assume that after an avalanche, for $f \equiv \phi^2 $ small enough,  the stress $\Sigma$ grows almost linearly in time starting from some value $\Sigma_0$. Thus the interevent time $t_i$ is defined by the
relation:
\begin{equation}\label{t_i_definition}
\dot \Gamma - \left(\Sigma_0+ \frac{t_i \dot \Gamma} {\tau}\right)\phi(t_i)^2 = 0
\end{equation}
i.e. at the time $t_i$ (computed from the end of the previous avalanche) when another avalanche starts. Then we can distinguish two limiting possible situations: 
a) when $\Sigma_0 \gg \frac{t_i \dot{\Gamma}}{\tau}$ and b) when $\Sigma_0 \ll \frac{t_i \dot{\Gamma}}{\tau}$. In case a) we can assume that small fluctuations of $\phi$ are enough to start a new avalanche. Then, the probability distribution of $t_i$ can be computed as the waiting time probability for a random walk, when $\phi$  exceeds the boundary $|\phi|= \sqrt{\dot \Gamma/\Sigma_0}$. In this case it is well known~\cite{Bray_Majumdar_Schehr_2013} that:
\begin{equation}\label{eq:A}
P_a[t_i]    \sim \frac{1}{t_i^{3/2}} .
\end{equation}
In the  other limiting case b) we can neglect 
$\Sigma_0$ in Eq.~\eqref{t_i_definition} and we obtain the condition 
\begin{equation}\label{t_i}
t_i \phi(t_i)^2 /\tau \equiv t_i f(t_i)/\tau \sim 1. 
\end{equation}
Note that both cases and in particular Eq.~\eqref{t_i} do not imply any correlation between avalanche size and interevent time (see also Fig.~\ref{fig:stress_timeseries}(b)). The value of $\phi$ in Eq.~\eqref{t_i} should correspond to the initial stage of the avalanche, the size of which depends on the magnitude of the stress and its subsequent dynamics. Using Eqs.~\eqref{eq:pdf_fluidity} and~\eqref{t_i}  we obtain 
\begin{equation} \label{eq:pdf_ti}
    P_b[t_i] = P_b[f]\left|\frac{df}{dt_i}\right|\sim \frac{1}{t_i^{3/2}[1+At_i^{-1}]^{1+\delta}}\sim\frac{1}{t_i^{1/2 - \delta}},
\end{equation}
in the limit of large $t_i$ and $A/t_i \gg1$. To summarize, our theoretical analysis allows us to identify the asymptotic behaviour of the $P[t_i]$ for small and large $t_i$.
Our findings and expectations are in agreement with the numerical results shown in Fig.~\ref{fig:pdf_ti_sizes}. Notice that, at large $t_i$, we observe an exponential cut-off due to finite size effects, as expected. This implies that the scaling~\eqref{eq:pdf_ti} can be obtained by fitting the large $t_i$ behavior of $P[t_i]$ with a Gamma distribution $G(t_i)$ given by $G(t_i) = C t_i^{-\gamma}\exp(-t_i /\kappa)$, with $C$ and $\kappa$ constants. Following this procedure, the scaling exponent $\gamma$ for different values of $\delta$ nicely agrees with the results predicted in Eq.~\eqref{eq:pdf_ti}. 
It is a remarkable fact that the results shown in Fig.~\ref{fig:pdf_ti_sizes} are in qualitative and almost quantitative agreement with what is observed in experimental observations and numerical simulations of granular systems and soft-glasses ~\cite{Corral_2006,Lherminier_Planet_Vehel_Simon_Vanel_Maloy_Ramos_2019,Kumar_Korkolis_Benzi_Denisov_Niemeijer_Schall_Toschi_Trampert_2020}. 
Importantly, the two scaling exponents $\alpha = 5/4+\delta/2$ for the size distribution $S$ and $\gamma = 1/2-\delta$ for the interevent time $t_i$ satisfy the relation: 
\begin{equation}\label{eq:scale_invariance}
\gamma = 3 - 2 \alpha.
\end{equation}
The validity of the above relation can also be directly checked by noticing that the variable $\chi \equiv 1/t_i^2$ has the same probability distribution as $S$ with extremely good agreement (see Sec.~\ref*{sec:ti_trick} of SM). Eq.~\eqref{eq:scale_invariance} was already deduced in Ref.~\cite{Benzi_Castaldi_Toschi_Trampert_2022} for a simplified model of turbulence displaying avalanche-like events. In particular, Eq.~\eqref{eq:scale_invariance} can be obtained  upon assuming that the ratio $X=S/t_i^2$ between energy dissipation $S$ and energy stored $\Gamma \Sigma \sim t_i^2$ is invariant by coarse-graining scale transformation $S_0 \rightarrow \lambda S_0$, i.e. by considering only avalanches larger than size $\lambda S_0$, where $S_0$ is the minimum observed avalanche size, and computing $t_i$ as the interevent time of avalanches larger than $\lambda S_0$. Following~\cite{Benzi_Castaldi_Toschi_Trampert_2022}, one can prove that using Eq.~\eqref{eq:scale_invariance} the probability distribution $P[t_i|\lambda S_0]$ maintains its functional form by increasing $\lambda$. More precisely, we expect a scale-dependent crossover: while the Gamma distribution remains invariant under increasing $\lambda$, the small-$t_i$ power-law regime is progressively suppressed, leading to a wider dominance of the Gamma behavior (See Sec.~\ref*{sec:coarsening} of SM for more details). Fig.~\ref{fig:0d_coarsening}(a) illustrates the probability distribution $P[t_i\vert{}\lambda S_0]$ for various values of $\lambda$. As $\lambda$ increases, the distribution asymptotically approaches a Gamma distribution with an exponent $\gamma = 1/2-\delta$. This agreement is further quantified in Fig.~\ref{fig:0d_coarsening}(b), which displays the ratio $R(t_i) = P[t_i \vert{} \lambda S_0]/ G(t_i)$. For comparison, Fig.~\ref{fig:0d_coarsening}(c) shows the corresponding ratio calculated using a pure exponential distribution ($\gamma=0$). The comparison underscores that the Gamma distribution accurately captures the behavior of $P[t_i \vert{} \lambda S_0]$, particularly in the large $\lambda S_0$ regime. Notably, for $\lambda S_0 = 0.1$, the data collapse under the Gamma fit (with $\gamma = 0.34$) spans nearly three orders of magnitude, whereas the exponential fit ($\gamma = 0$) holds for barely one. In Sec.~\ref*{sec:space_dependent_statistics} of SM we show that the same analysis can be done for the spatial dependent model given by Eqs.~\eqref{eq:maxwell_model} and~\eqref{eq:fluidity_eq}. We note that the invariance of the probability distribution is lost, if instead of considering a minimum size $\lambda S_0$, we consider the set of avalanches for a changing threshold $M_t$. This implies that these are two different methods to coarse-grain (see Sec.~\ref*{sec:changing_definition} of SM for more details).

\begin{figure}[]
    \centering
    \includegraphics[width=0.9\linewidth]{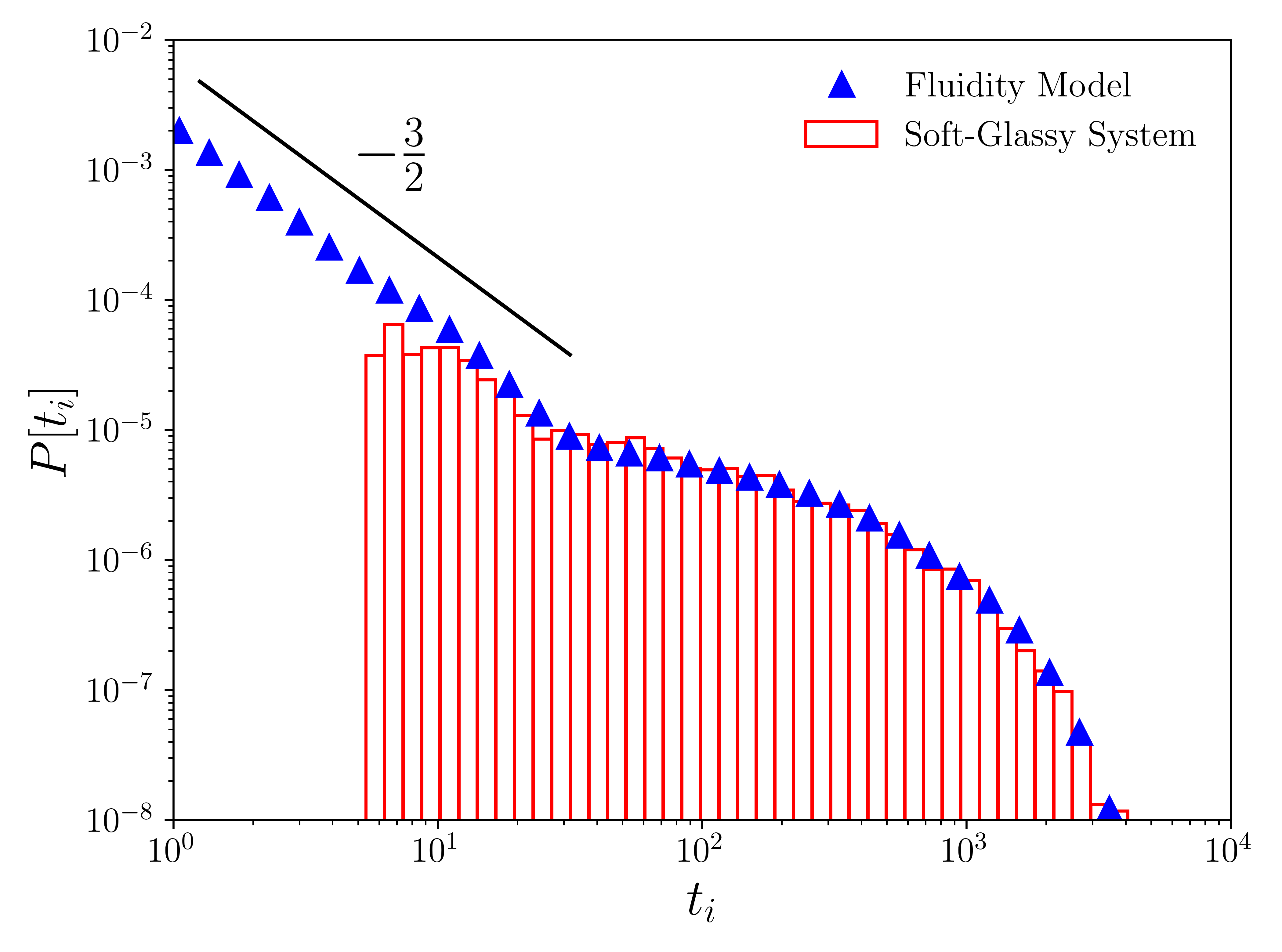}
    \caption{
    Comparison of the probability distribution $P[t_i]$ of the interevent time between the numerical result obtained from the fluidity model, Eqs.~\eqref{eq:maxwell_model} and~\eqref{eq:0d_fluidity}  (blue triangles),  against results from a numerical simulation of a soft-glassy system (red boxes).}
    \label{fig:ti_lbm}
\end{figure}

In many different amorphous systems, like soft-glasses or granular systems, the values of exponents $\alpha$ and $\gamma$ are close to $1.33$ and $0.34$
respectively ~\cite{Davidsen_Kwiatek_2013,Kumar_Korkolis_Benzi_Denisov_Niemeijer_Schall_Toschi_Trampert_2020}.  In particular in Ref.~\cite{Kumar_Korkolis_Benzi_Denisov_Niemeijer_Schall_Toschi_Trampert_2020} a direct comparison between numerical simulations and experimental data shows the same scaling exponents $\alpha$ and $\gamma$. In our model, these exponents can be obtained for  $\delta = 0.16$, as shown in Fig.~\ref{fig:pdf_ti_sizes}.
The numerical simulations in Ref.~\cite{Kumar_Korkolis_Benzi_Denisov_Niemeijer_Schall_Toschi_Trampert_2020} were performed for a spatially resolved two dimensional soft-glassy system. In Fig.~\ref{fig:ti_lbm} 
we show the probability distribution of the interevent time $t_i$ obtained from the soft-glassy system compared to the one obtained by our fluidity model using  Eqs.~\eqref{eq:maxwell_model} and~\eqref{eq:0d_fluidity} with $\delta = 0.16$.
The time of the numerical soft-glassy simulation has been rescaled for comparison. We observe that not only the long-time behavior agrees with the simple model, but also the short-time behavior agrees with the simplified model, although in the numerical soft-glassy system very short timescales were not available. The excellent agreement shown in Fig.~\ref{fig:ti_lbm} strongly supports the validity of our 
model. 
%%%%%%%%%%%%%%%%%%%%%%%%%%%%%%%%%%%%%%%%%%%%%%%%%%%%%%%%%
%%%%%%%%%%%%%%%%%%%%%%%%%%%%%%%%%%%%%%%%%%%%%%%%%%%%%%%%%%%%%%%%%%%%%%%

In summary, we have shown that our model described by Eqs.~\eqref{eq:maxwell_model} and~\eqref{eq:fluidity_eq} is able to capture many features observed in avalanche dynamics. Most importantly we are able to predict the interevent time distribution and its scale invariance under coarse-graining in the avalanche size 
with an outstanding comparison against existing data. Notably, although the model is formulated in terms of multiple parameters ($\epsilon_0, \epsilon_1, \tau, \xi, n, \dots$), the scaling properties of both the avalanche size distribution and the interevent times depend solely on the single parameter $\delta$. Let us recall that $\delta$ represents the ratio of two distinct timescales and dictates the magnitude of the correlation scale, as established in Eq.~\eqref{eq:lc}. This unique dependence highlights the inherent robustness of our approach. To the best of our knowledge, this model is the first to derive -- rather than postulate -- the Gamma distribution of interevent times observed across diverse physical systems~\cite{Davidsen_Kwiatek_2013,Corral_2006,Kumar_Korkolis_Benzi_Denisov_Niemeijer_Schall_Toschi_Trampert_2020}, achieving this from just two fundamental ingredients: a Maxwell element and a dynamical equation for the relaxation time.
\\
\emph{Acknowledgments.} Funding for this research was provided by DeepNL: A digital twin for modeling and forecasting induced seismicity (2023/ENW/01616324 17780). Support from INFN through project FIELDTURB is also acknowledged.
\\
%\emph{Data Availability.} The data and code that support the findings in this manuscript are openly available on \textit{Zenodo} \cite{dataavail}.
\\
\bibliography{references}% Produces the bibliography via BibTeX.

\renewcommand{\theequation}{S\arabic{equation}}
\renewcommand{\thefigure}{S\arabic{figure}}
\renewcommand{\thetable}{S\arabic{table}}
\setcounter{equation}{0}
\setcounter{figure}{0}
\setcounter{table}{0}
\renewcommand{\thesection}{\Roman{section}}
\renewcommand{\thesubsection}{\Roman{subsection}}
\section*{Supplementary Material}
%%%%%%%%%%%%%%%%%%%
\subsection{Model equations}\label{sec:model_equations}
%%%%%%%%%%%%%%%%%

The model used in this paper has been developed in~\cite{Benzi_Sbragaglia_Bernaschi_Succi_Toschi_2016, Benzi_Divoux_Barentin_Manneville_Sbragaglia_Toschi_2019, Benzi_Divoux_Barentin_Manneville_Sbragaglia_Toschi_2021, Benzi_Divoux_Barentin_Manneville_Sbragaglia_Toschi_2023}.  The starting point is the stationary fluidity equation introduced by~\cite{Bocquet_Colin_Ajdari_2009} 
\begin{equation}\label{b1}
\xi^2 \Delta f + m(\Sigma)f -f^{3/2} = 0,
\end{equation}
written using the same notation as in the main text. Eq.~\eqref{b1} has been derived in~\cite{Bocquet_Colin_Ajdari_2009} in a suitable coarse grained limit from the elastoplastic model~\cite{Nicolas_Ferrero_Martens_Barrat_2018} and it describes how the rheological behaviour of a plastic flow depends on the system size $\xi/L$, in extremely good agreement with experiments~\cite{Goyon_Colin_Ovarlez_Ajdari_Bocquet_2008}. Following~\cite{Bocquet_Colin_Ajdari_2009}, Eq.~\eqref{b1} can be derived from a 'free energy' functional $F[\phi]$, which includes a bulk potential and nonlocal terms. In a series of papers~\cite{Benzi_Sbragaglia_Bernaschi_Succi_Toschi_2016, Benzi_Divoux_Barentin_Manneville_Sbragaglia_Toschi_2019, Benzi_Divoux_Barentin_Manneville_Sbragaglia_Toschi_2021},  Eq.~\eqref{b1} has been generalized to take into account the non-stationary evolution in shear and stress start-up experiments. This was done by assuming that the time dynamics follows $\frac{\partial \phi}{\partial t} = - \frac{\delta F}{\delta \phi}$.
In its final version the full dynamical model reads:
\begin{eqnarray}
    \label{b3}
    \frac{d\Sigma}{dt} &=& \frac{1}{\tau}[\dot \Gamma - \langle f \rangle \Sigma]\\
    \label{b2}
    \frac{\partial f}{\partial t} &=& {\dot \Gamma} f [ \xi^2 \Delta f + m(\Sigma)f -f^{3/2} ].
\end{eqnarray}
The model defined in Eqs.~\eqref{b3}-\eqref{b2} has been used to compute analytically the scaling behavior of the fluidization time as a function of the imposed shear $\dot \Gamma$ or stress and the scaling of the stress overshoot as a function of the imposed shear $\dot \Gamma$. Importantly, although the limit $\xi \rightarrow 0$ is singular, all the predictions depend only on the Herschel-Bulkley exponent $n$ which appears in the definition of $m(\Sigma)$. \\
An important physical choice concerns the boundary conditions. As discussed in Ref ~\cite{Benzi_Divoux_Barentin_Manneville_Sbragaglia_Toschi_2021, Benzi_Divoux_Barentin_Manneville_Sbragaglia_Toschi_2023}, we fix $\partial_y f=0$ at $y=0$ while for $y=L$ we have two possible choices: 1) $f(L)=m^2(\Sigma)$ and 2) $\partial_y f(L)=0$. Choice 1) produces a fluidization layer near $y=L$ of width $l_b$, which, for small enough $\epsilon_0$ ~\cite{Benzi_Sbragaglia_Bernaschi_Succi_Toschi_2016}, can eventually increase with time reaching $y=0$. The choice 2) is assumed to be consistent with very stiff materials and  with a brittle-like transition and avalanche dynamics.\\
In this paper we work at constant (small) $\dot \Gamma$. Then by the change of variables $ \tilde t = t \dot \Gamma$ and $\tilde \tau = \tau \dot \Gamma$, the model can be rewritten as:
\begin{eqnarray}
    \label{b5}
    \frac{d\Sigma}{d \tilde t} &=& \frac{1}{\tilde \tau}[\dot \Gamma - \langle f \rangle \Sigma] \notag \\
    \label{b4}
    \frac{\partial f}{\partial \tilde t} &=&  f [ \xi^2 \Delta f + m(\Sigma)f -f^{3/2} ] \notag.
\end{eqnarray}
Finally by renaming $t$ and $\tau$ and using $f = \phi^2$ the model becomes:
\begin{eqnarray}
    \label{b7}
    \frac{d\Sigma}{dt} &=& \frac{1}{\tau}[\dot \Gamma - \langle \phi^2 \rangle \Sigma]\\
    \label{b6}
    \frac{\partial \phi}{\partial t} &=&   2\xi^2 \phi^2 \Delta \phi  +2\xi^2 \phi(\nabla\phi)^2 + m(\Sigma) \phi^3 - \phi^3|\phi| 
\end{eqnarray}
In Ref.~\cite{Benzi_Sbragaglia_Bernaschi_Succi_Toschi_2016}, Eq.~\eqref{b6} has been investigated in detail. By assuming $\dot \Gamma = const$ or $\Sigma = const$ (in dynamical equilibrium dictated by Eq.~\eqref{b7}, it can be shown that Eq.~\eqref{b6} exhibits localized stable solutions corresponding to a shear band assuming the proper boundary conditions. To model mechanical noise due to heterogeneities, a white noise term, $\sqrt{\epsilon_0} \eta_1(y,t)$,  was added to the r.h.s of Eq.~\eqref{b6}. The next step in the development was to add a multiplicative noise term, $\sqrt{\epsilon_1 \langle \phi^2 \rangle} \eta_2(t)$, which is space-independent. This fluctuation can enhance the stochastic perturbations with a positive feedback loop, causing avalanche like behavior \cite{Benzi_Sbragaglia_Bernaschi_Succi_Toschi_2016}. Then the full equation for the fluidity reads
\begin{eqnarray}
        \frac{\partial \phi}{\partial t} =&   2\xi^2 \phi^2 \Delta \phi  +2\xi^2 \phi(\nabla\phi)^2 + m(\Sigma) \phi^3 - \phi^3|\phi| \notag \\
        &+ \sqrt{\epsilon_0} \eta_1(y,t) + \sqrt{\epsilon_1 \langle \phi^2 \rangle} \eta_2(t).\label{eq:sup_fluidity}
\end{eqnarray}
% scale invariance
One property of Eqs.~\eqref{b5}-\eqref{eq:sup_fluidity}, which is used in the main text to find scaling relations, is that they are invariant under the following transformation
\begin{align*}
\phi &\rightarrow \lambda \phi & m &\rightarrow \lambda m \\
\Sigma &\rightarrow \lambda^2 \Sigma &t &\rightarrow \lambda^{-3} t \\
\dot{\Gamma} &\rightarrow \lambda^4 \dot \Gamma & \epsilon_1 &\rightarrow \lambda^3 \epsilon_1 \\
\epsilon_0 &\rightarrow \lambda^5 \epsilon_0 &\tau &\rightarrow \lambda^{-1} \tau.
\end{align*}
%%%%%%%%%%%%%%%%%%%%%%%%%%%%
\subsection{Self-consistent field approximation}\label{sec:self-consistent-approx}
%%%%%%%%%%%%%%%%%%%%%%%%%%%
One can rewrite Eq.~\eqref{eq:sup_fluidity} in the following way:
\begin{equation}
    \label{eq_F}
    \partial_t \phi = - \frac{\delta F}{\delta \phi}+ \sqrt{\epsilon_0} \eta_1(y,t) + \sqrt{\epsilon_1 \langle \phi^2 \rangle} \eta_2(t) \notag
\end{equation}
where
\begin{equation}
    F[\phi] =  \int dy \left[\xi^2 \phi^2 (\nabla \phi)^2 -\frac{1}{4} m \phi^4+\frac{1}{5} \phi^4|\phi| \right] \notag.
\end{equation}
The self-consistent approximation is obtained by using Hartree method \cite{Benzi_Sbragaglia_Bernaschi_Succi_Toschi_2016}. In this case we have:
\begin{equation}
    \int dy \left[\xi^2 \phi^2 (\nabla \phi)^2\right] \approx\int dy  \left[ \xi^2\langle \phi^2 \rangle (\nabla \phi)^2 + \xi^2 \phi^2 \langle (\nabla \phi)^2 \rangle  \right]. \notag
\end{equation}

Next, by defining the following renormalized parameters
\begin{eqnarray}
    \label{eq:def_D}
    D &=& 2\xi^2 \langle \phi^2 \rangle \\
    \label{eq:def_R}
    R &=& 2 \xi^2 \langle (\nabla \phi)^2 \rangle,
\end{eqnarray}
we can rewrite Eq.~\eqref{eq:sup_fluidity} to 
\begin{align}
     \frac{\partial \phi}{\partial t} =&  D \Delta\phi -R \phi + m(\Sigma)\phi^3 - \phi^3 |\phi|  \notag \\
     &+ \sqrt{\epsilon_0} \eta_1(y,t) 
+ \sqrt{\epsilon_1 \langle \phi^2 \rangle} \eta_2(t). \label{eq:sup_renormalized}
\end{align}

To derive the correlation length in one dimension for Eq.~\eqref{eq:sup_renormalized}, we start from the linearized equation
\begin{equation}
    \partial_t \phi = D \partial_y^2\phi - R \phi + \sqrt{\epsilon_0} \eta_1(y,t) + \sqrt{\epsilon_1 \langle \phi^2\rangle} \eta_2(t) \notag.
\end{equation}
After Fourier transforming with
\begin{equation}
   \phi_k(t) = \frac{1}{\sqrt{2\pi}}\int e^{iky} \phi(y,t) dy \notag,
\end{equation}
we obtain a set of equations for the Fourier modes
\begin{equation} \label{eq_sup:linear_k}
    \partial_t \phi_k = -(D k^2+R) \phi_k + \sqrt{\epsilon_0} \eta_1(k,t) + \sqrt{2\pi \epsilon_1 \langle\phi^2\rangle} \eta_2(t) \delta(k),
\end{equation}
where we have defined $\eta_1(k,t) = \frac{1}{\sqrt{2\pi}} \int e^{iky} \eta_1(y,t) dy$. Eq.~\eqref{eq_sup:linear_k} implies that the $k$ modes decouple and that for every $k$ mode we have an Ornstein-Uhlenbeck process, for which in the steady state we find 
\begin{eqnarray}
    \langle \phi_k \phi_{-k} \rangle = \frac{1}{2}\frac{\epsilon_0}{Dk^2 +R} + \frac{\pi \epsilon_1 \langle \phi^2 \rangle \delta(k)}{D k^2 + R},
\end{eqnarray}
from which we can compute the average fluidity in real space via Parseval's identity, $\langle \phi^2\rangle = \frac{1}{2\pi} \int dk \langle\phi_k \phi_{-k}\rangle$, to get
\begin{equation}
    \langle \phi^2\rangle 
    = \frac{1}{4\pi} \int\frac{\epsilon_0 dk}{Dk^2 + R} + \frac{1}{2} \frac{\epsilon_1 \langle \phi^2 \rangle}{R} \notag.
\end{equation}
Plugging this result into Eq.~\eqref{eq:def_D} gives us
\begin{equation}\label{eq:D}
    D =  \frac{\epsilon_0 \xi^2}{2\pi} \int \frac{dk}{D k^2 + R}+ \frac{\epsilon_1 D}{2 R}. 
\end{equation}
Using Eq.~\eqref{eq:def_R}, we find
\begin{equation}
    R = \frac{\xi^2}{\pi} \int dk \langle k^2 \phi_k \phi_{-k} \rangle =\frac{\epsilon_0 \xi^2}{2\pi} \int \frac{k^2 dk}{Dk^2 + R} \label{eq:R}.
\end{equation}
Multiplying Eqs.~\eqref{eq:D} and~\eqref{eq:R} by $R$ and $D$, respectively, and summing both results, leads to
\begin{eqnarray}\label{eq:2DR}
2DR = \frac{\epsilon_0 \xi^2}{2 \pi} \int dk + \frac{\epsilon_1  D}{2}. \notag
\end{eqnarray}
To regularize the integral, we impose a cutoff $k_M$, which is the inverse of the smallest length scale in the system. This allows us to perform the integral and we find
\begin{equation}\label{eq:DR}
    DR = \frac{\epsilon_0 \xi^2 k_M}{2\pi(1-\frac{\epsilon_1}{4R})}.
\end{equation}
Next, we can explicitly perform the integration in Eq.~\eqref{eq:D} 
\begin{equation} \label{eq:D1}
D = \frac{\epsilon_0 \xi^2}{\pi} \frac{\tan^{-1}({\frac{\sqrt{D}}{\sqrt{R}}k_M})}{\sqrt{DR}} + \frac{\epsilon_1 D}{2 R} 
    =\frac{\epsilon_0 \xi^2}{2\sqrt{DR}} + \frac{\epsilon_1 D}{2R},
\end{equation}
where we have used $k_M \gg 1$. Using the result from Eq.~\eqref{eq:DR} for the first term on the right hand-side of Eq.~\eqref{eq:D1} and solving for $D$, we find
\begin{eqnarray}
    D = \frac{\sqrt{\pi \epsilon_0 \xi^2}}{\sqrt{2k_M}} \frac{\sqrt{1-\frac{\epsilon_1}{4R}}}{1- \frac{\epsilon_1}{2R}} = \frac{D_0}{1-\frac{3}{8}\frac{\epsilon_1}{R_0}} + O(\epsilon_1^2) \label{eq:D_e1},
\end{eqnarray}
where $D_0 =\frac{\xi\sqrt{\pi \epsilon_0 }}{\sqrt{2k_M}}$ and $R_0$ is $R$ for $\epsilon_1 = 0$. To find an equivalent expression for $R$, we combine the results of Eqs.~\eqref{eq:DR} and~\eqref{eq:D_e1},
\begin{equation}
    R = \frac{\epsilon_0 \xi^2 k_M}{2\pi D_0} \frac{1-\frac{3}{8} \frac{\epsilon_1}{R}}{1-\frac{\epsilon_1}{4R}} + O(\epsilon_1^2) = R_0 \left(1 - \frac{\epsilon_1}{8R_0}\right) + O(\epsilon_1^2),
\end{equation}
with $R_0 = \frac{\xi \sqrt{\epsilon_0} k_M^{3/2}}{\sqrt{2} \pi^{3/2}}$. Then we can find the correlation length in the system to first order in $\epsilon_1$ as
\begin{equation}
    l_c = \sqrt{\frac{D}{R}} \sim  \frac{l_0}{\sqrt{1-1/(2\delta)}} +O(\epsilon_1^2),
\end{equation}
were
$$
\delta = \frac{R_0}{\epsilon_1}
$$ 
and 
$$
l_0 = \sqrt{\frac{D_0}{R_0}} \sim \frac{1}{k_M}.
$$ 
Thus we find that the correlation length in the system increases with $\epsilon_1$ and that it becomes system size when $\epsilon_1 \sim 2 R_0$. If the correlation length has become system size we can approximate Eq.~\eqref{eq:sup_renormalized} without space dependency as
\begin{align}\label{eq:sup_fluidity_0d}
     \frac{d \phi}{dt} =&  -R \phi + m(\Sigma)\phi^3 - \phi^3 |\phi|
     + \sqrt{\epsilon_0 + \epsilon_1 \phi^2 } \eta(t). 
\end{align}
%%%%%%%%%%%%%%%%%%%%%%%%
\subsection{Probability distributions}\label{sec:probability_distributions}
\begin{figure}[h!]
    \centering
    \includegraphics[width=0.9\linewidth]{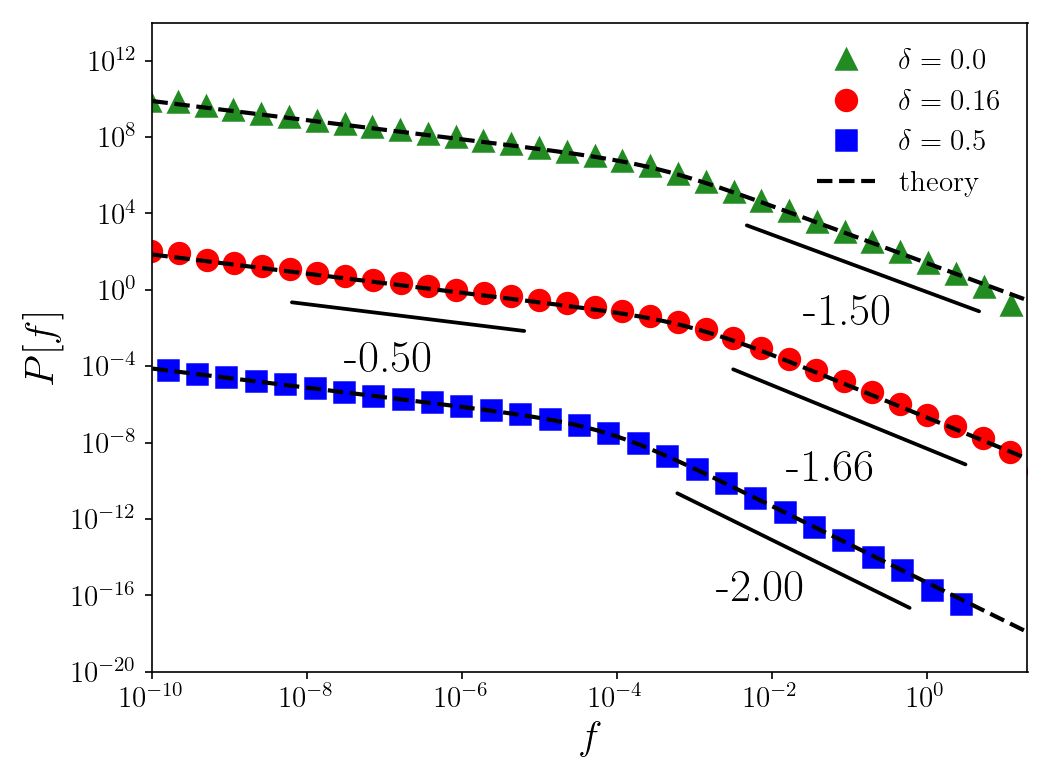}
    \caption{
    Probability density $P[f]$ for the simplified fluidity model described by Eqs.~\eqref{eq:maxwell_model} and~\eqref{eq:0d_fluidity} for different values of $\delta$.  Black dashed lines are the analytical prediction given by Eq.~\eqref{eq_sup:pdf_fluidity}. Results obtained with the same parameters as in Fig.~\ref{fig:pdf_ti_sizes} with $\epsilon_0 = 10^{-6}, \tau = 0.1$, for different values of $\delta$.}
    \label{fig:fluidity_pdf}
\end{figure}

%%%%%%%%%%%%%%%%%%%%%%%%%%%%%%%%%%
To obtain an analytical prediction for the probability distribution $P[\phi,t]$, we can use the Fokker-Planck equation related to the SDE given in Eq.~\eqref{eq:sup_fluidity_0d}, which is given by
\begin{equation}
    \frac{\partial P[\phi,t]}{\partial t}
    = -\frac{\partial}{\partial \phi}\big[A(\phi)\,P[\phi,t]\big]
    + \frac{\partial^2}{\partial \phi^2}\big[D(\phi)\,P[\phi,t]\big] \notag,
\end{equation}
where $A(\phi) = -R \phi + m(\Sigma) \phi^3 - \phi^3 |\phi|$ and $D(\phi) = (\epsilon_0 + \epsilon_1 \phi^2)/2$. To find an analytical solution, we assume that the stress is constant and equal to $\Sigma = \Sigma_c$, which is generally not true as can be seen in Fig.~\ref{fig:stress_timeseries}, but we will show that it still captures the correct statistics.
Using this assumption, one can show that the stationary Fokker-Planck equation (i.e. $\partial_t P[\phi,t] = 0$), is solved by the probability distribution 
\begin{equation}
    P[\phi] =\frac{C}{1 + A \phi^2} \exp \left(\frac{2}{\epsilon_0} \int^\phi \frac{-R \phi' + m \phi'^3 - \phi'^3 |\phi'|}{1 + A \phi'^2} d\phi' \right) \notag
\end{equation}

with $A=\frac{\epsilon_1}{\epsilon_0}$ and $C$ being a normalization constant. 
The above equation can also be written in the limit of small $\epsilon_0$ as:
\begin{equation}
    P[\phi] =  \frac{C}{[1+A \phi^2]^{1+\delta}} \exp \left( \frac{m \phi^2-\frac{2}{3}\phi^2|\phi|}{\epsilon_1}+ O(\epsilon_0) 
    \right) \notag
\end{equation}
To obtain the probability distribution in terms of $f = \phi^2$, we do a change of variables, obtaining
\begin{equation} \label{eq_sup:pdf_fluidity}
    P[f] = \frac{\bar{C}}{f^{1/2}(1+Af)^{1+\delta}} \exp\left(N(f)\right),
\end{equation}
with $N(f) = \frac{m f - \frac{2}{3} f^{3/2}}{\epsilon_1} + O(\epsilon_0)$, $\bar{C}$ is a normalization constant and where we have defined $\delta = \frac{R}{\epsilon_1}$.
We can distinguish two different scaling regimes, which can be fully characterized by the two parameters $A, \delta$. For $f \ll 1/A$, we find that $P[f]\sim \frac{1}{f^{1/2}}$ which is independent of the parameters used. For $f\gg 1/A$, we find, in the limit of small $\epsilon_0$, $P[f] \sim \frac{1}{f^{3/2 + \delta}}\exp \left(\frac{m f- \frac{2}{3}f^{3/2}}{\epsilon_1}\right)$. Thus we have a power-law followed by an exponential cutoff. This exponential cutoff only dominates over the power law for large values of $f$, where the system spends almost no time. This is numerically confirmed in Fig.~\ref{fig:fluidity_pdf} for $A= 10^{6}$ and different values of $\delta$.

\subsection{Correlation between sizes and interevent time} \label{sec:correlation}
In the main text Fig.~\ref{fig:stress_timeseries}(b), we show the joint probability distribution of $t_i$ and $S$ for the simplified model given by Eqs.~\eqref{eq:maxwell_model} and~\eqref{eq:0d_fluidity} for $\delta = 0.16$. From Fig.~\ref{fig:stress_timeseries}(b) we see that there is no clear correlation present. Next, we do a quantitative check, by calculating the Pearson correlation coefficient. For the set of sizes $S = \{S_1, S_2, \dots, S_n\}$ and the corresponding interevent times
$t_i = \{t_{i,1}, t_{i,2}, \dots, t_{i,n}\}$, the Pearson correlation coefficient is
\begin{equation}
r = \frac{\sum_{k=1}^{n} (S_k - \bar{S})(t_{i,k} - \bar{t}_i)}
         {\sqrt{\sum_{k=1}^{n} (S_k - \bar{S})^2}\sqrt{\sum_{k=1}^{n} (t_{i,k} - \bar{t}_i)^2}},
\end{equation}
where $\bar S, \bar t_i$ are the average size and interevent time.
We find $r=3\times10^{-3}$, $\bar{t}_i = 3\times 10^{-1}$, $\bar{S} = 9\times 10^{-3}$ indicating that there is no clear correlation between the interevent times and sizes. 

\subsection{Coarsening the signal}\label{sec:coarsening}
\begin{figure}
    \centering
    \includegraphics[width=0.99\linewidth]{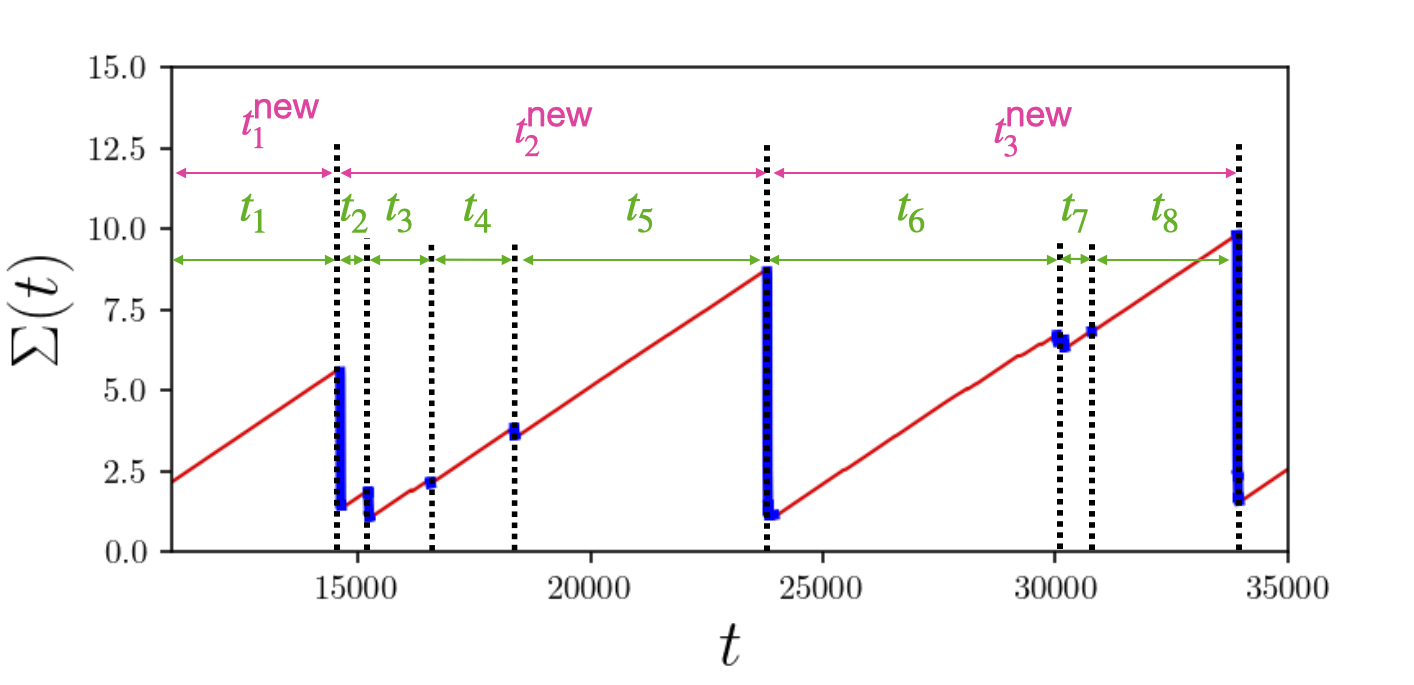}
    \caption{The stress signal $\Sigma(t)$ (red line) as function of time. The blue lines represent intervals where $d\Sigma/dt < 0$. $t_1 - t_8$ are the interevent times with no threshold on the avalanche size. $t_1^{\text{new}}- t_3^\text{new}$ are the interevent times after neglecting small avalanches. }
    \label{fig:coarsening_signal}
\end{figure}

In Fig.~\ref{fig:coarsening_signal} we give a schematic overview of what happens to the interevent time when a threshold on the avalanche size is applied. The shorter interevent times $t_2, t_3, t_4$ and $t_7$ get combined into longer interevent times.

\begin{figure}[h!]
    \centering
    \includegraphics[width=0.99\linewidth]{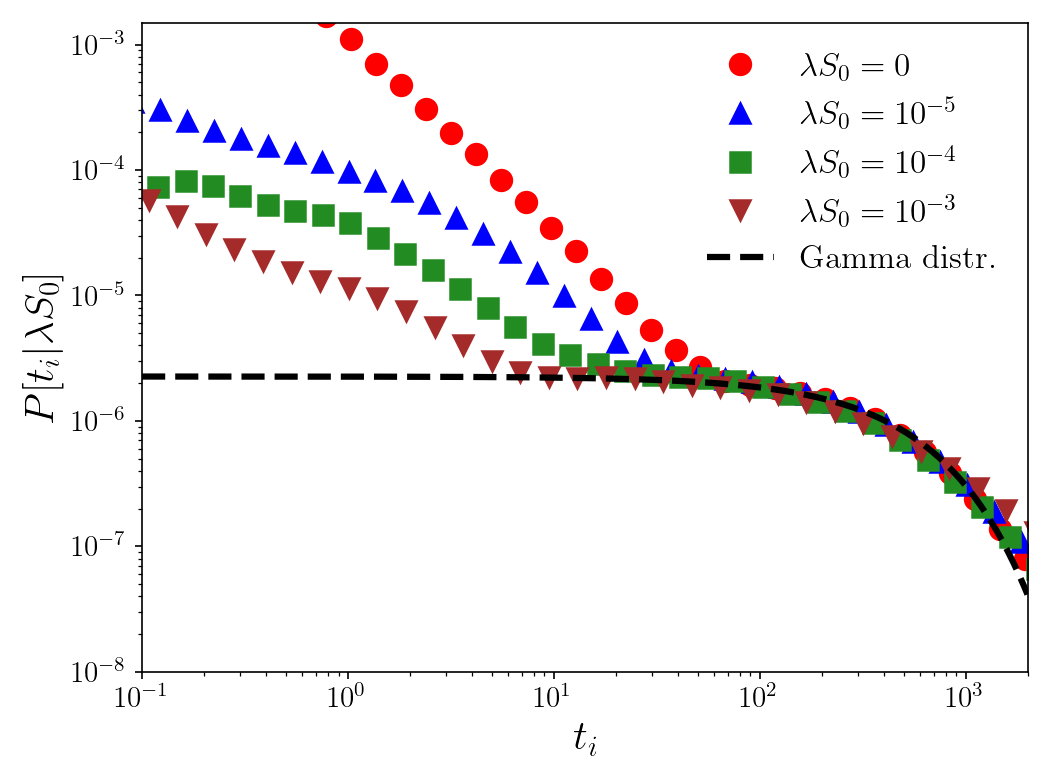}
    \caption{Probability density $P[t_i|\lambda S_0]$ for $\delta = 1/2$ and different values of $\lambda S_0$, for the model described by Eqs.~\eqref{eq:maxwell_model} and~\eqref{eq:0d_fluidity}. The black dashed line is the Gamma distribution with $\gamma = 0.0$ (exponential). Same parameters used as in Fig.~\ref{fig:pdf_ti_sizes}, with $R=0.05$.}
    \label{fig:0d_ti_delta_0.5}
\end{figure}

In the main part of the text, we have seen that the analytical prediction given by Eq.~\eqref{eq:pdf_ti} matches for $\delta = 0.16$. Here we show that the interevent time distribution after coarsening is consistent with $\delta = 1/2$. From Eq.~\eqref{eq:pdf_ti}, the interevent time distribution should then be exponential, this is confirmed in Fig.~\ref{fig:0d_ti_delta_0.5}.

\subsection{Critical value $\delta$}\label{sec:critical_signal}
We can give a simple argument as to why $\delta = 1/2$ corresponds to a critical value. Starting from Eq.~\eqref{eq:0d_fluidity}, we only consider the linear parts and assume $\sigma = 0$. In Itô form, we write as
\begin{equation}
    d\phi = -R \phi dt + \sqrt{\epsilon_0 + \epsilon_1 \phi^2}dW, \notag
\end{equation}
where $dW$ is the increment of a Wiener process. Next, to transform into $f = \phi^2$, we can use Itô's lemma 
$$
df  = \left(-R \phi \frac{df}{d\phi}  +  \frac{(\epsilon_0+\epsilon_1 \phi^2)}{2} \frac{d^2f}{d \phi^2} \right) dt + \sqrt{\epsilon_0 + \epsilon_1 \phi^2} \frac{d f}{d \phi} dW
$$
and obtain
\begin{equation}
df=\left(\epsilon_1f(1-2\delta)+\epsilon_0\right) dt + \sqrt{4(\epsilon_0 f + \epsilon_1 f^2)}dW.
\end{equation}
Now we see that if $\delta < \frac{1}{2}$ $f$ will grow in time, whereas for $\delta > 1/2$ $f$ will decrease to a small quantity proportional to $\epsilon_0$. Thus the value of $\delta = 1/2$ is exactly the value where the large fluidity fluctuations start happening.

\subsection{Relating time and sizes}\label{sec:ti_trick}
\begin{figure}[h!]
    \centering
    \includegraphics[width=0.9\linewidth]{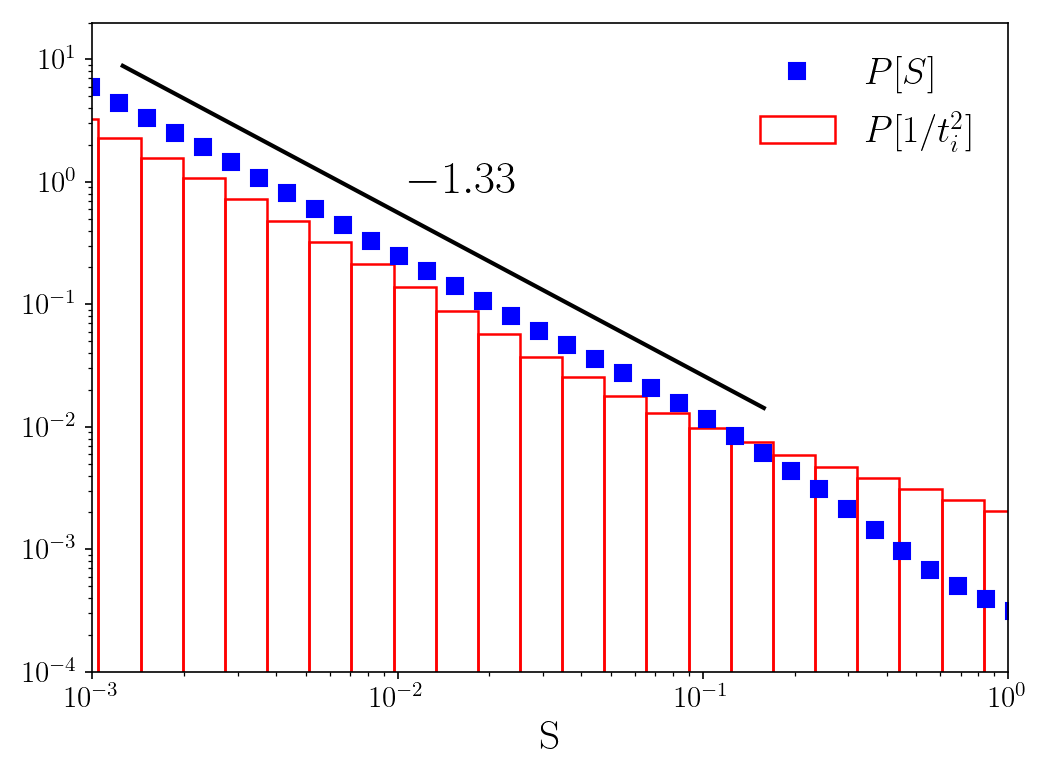}
    \caption{A quantitative check for our theoretical approach. According to the theory of scale invariance, Eq.~\eqref{eq:scale_invariance}, the probability density $P[1/t_i^2]$ should show the same scaling as the probability distribution $P[S]$. The same slope of $-1.33$ can be observed between $t_i \sim 10^{-3}$ and $10^{-1}$, agreeing with the theory.}
    \label{fig:ti_trick}
\end{figure}

In Ref \cite{Benzi_Divoux_Barentin_Manneville_Sbragaglia_Toschi_2021} a theoretical framework is given to link the statistics of the avalanche sizes with the interevent time. Using the relation given by Eq.~\eqref{eq:scale_invariance}, one can show that the probability distribution $P[1/t_i^2]$ obeys the same scaling as $P[S] = S^{-\alpha}$. In Fig.~\ref{fig:ti_trick} this is numerically confirmed for $\alpha = 1.33$, where we see that the scaling is the same on almost two orders of magnitude. This gives us a non-trivial check between the interevent times and the sizes. 

\subsection{Statistics of space-dependent model}\label{sec:space_dependent_statistics}
\begin{figure}[h!]
    \centering
    \includegraphics[width=0.99\linewidth]{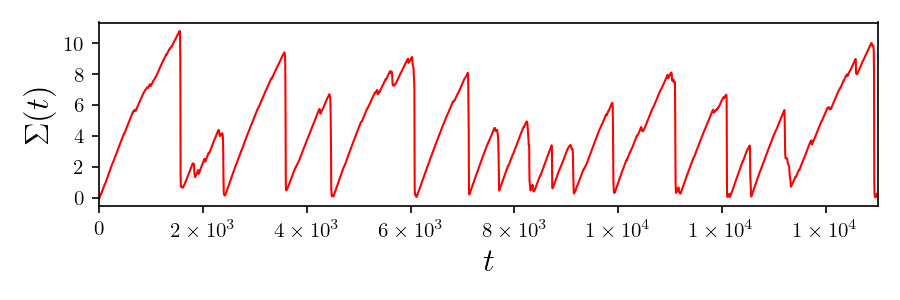}
    \caption{The stress signal $\Sigma(t)$ as function of time for the space dependent fluidity model given by Eqs.~\eqref{eq:maxwell_model} and~\eqref{eq:fluidity_eq}. Numerical results obtained with $\dot \Gamma = 0.01, \tau =1.0, \epsilon_0 =9.8\cdot 10^{-8}, \epsilon_1 = 0.1, \xi = 0.02$.}
    \label{fig:stress_signal_1d}
\end{figure}

\begin{figure}[h!]
    \centering
    \includegraphics[width=0.89\linewidth]{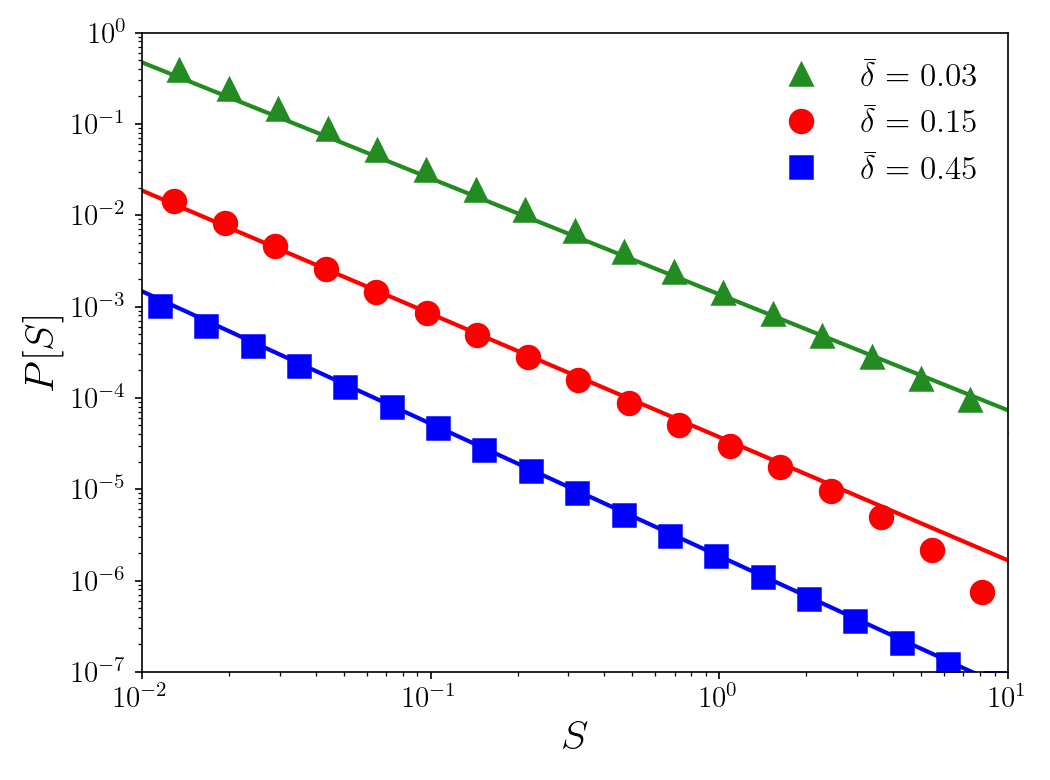}
    \caption{The probability density $P[S]$ for the avalanche sizes for the model described by Eqs.~\eqref{eq:maxwell_model} and~\eqref{eq:fluidity_eq}. Here we use a time averaged value  $\bar \delta$, to control the system. The prediction for the avalanche sizes in Eq.~\eqref{eq:pdf_sizes} matches the numerical results for the space dependent model. Same parameters used as in Fig. \ref{fig_sup:1d_ti}.}
    \label{fig:pdf_1d_sizes}
\end{figure}

\begin{figure}[h!]
    \centering
    \includegraphics[width=0.89\linewidth]{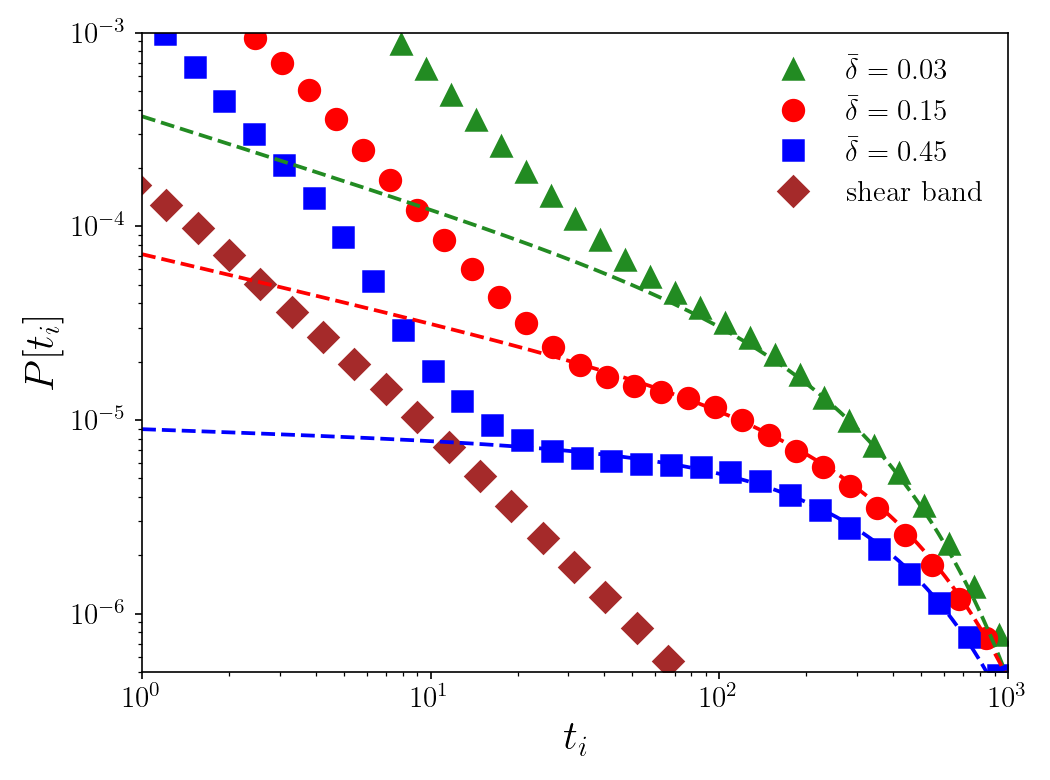}
    \caption{The probability density $P[t_i]$ of the interevent time for the model described by Eqs. \eqref{eq:maxwell_model} and \eqref{eq:fluidity_eq} for the case with a shear band (bottom) and the case without a shear band (top three). In the latter case, a clear long-tail is visible which can be well fitted by a Gamma distribution (dashed lines). Numerical results obtained with $\dot \Gamma = 0.01, \tau =0.1, \epsilon_0 =2 \cdot 10^{-10}, \epsilon_1 = 0.1$, $\xi = 0.1$ (green), $\xi = 0.4$ (blue), $\xi = 1.0$ (red). }
    \label{fig_sup:1d_ti}
\end{figure}

We show that we can obtain the same results for the avalanche statistics in the spatial model given by Eq.~\eqref{eq:fluidity_eq} compared to the simplified model given by Eq.~\eqref{eq:0d_fluidity}. To model the dynamics of a brittle material, we use the boundary condition $\partial_y \phi |_L = 0$, which corresponds to material having no shear band or $l_b = 0$. In Fig.~\ref{fig:stress_signal_1d}, we see that the stress drops still show a short and long timescale. For the simplified model we have seen that the only relevant parameter controlling the avalanche statistics is $\delta$.  For the space dependent model, we have defined $R(t) = 2\xi^2\langle(\partial_y\phi(y,t))^2\rangle$. Since in this case the parameter $\delta(t) = R(t)/ \epsilon_1$, depends on time, we use the time averaged value $\bar \delta $ as a way to control the system. Using the time averaged value $\bar \delta$, we show in Fig.~\ref{fig:pdf_1d_sizes} that the prediction for the avalanche size, given by Eq.~\eqref{eq:pdf_sizes} still holds. Also, we show in Fig.~\ref{fig_sup:1d_ti} that the long tail in the interevent time statistics can still be well captured by a Gamma distribution (top three curves).

\begin{figure}[h!]
    \centering
    \includegraphics[width=0.89\linewidth]{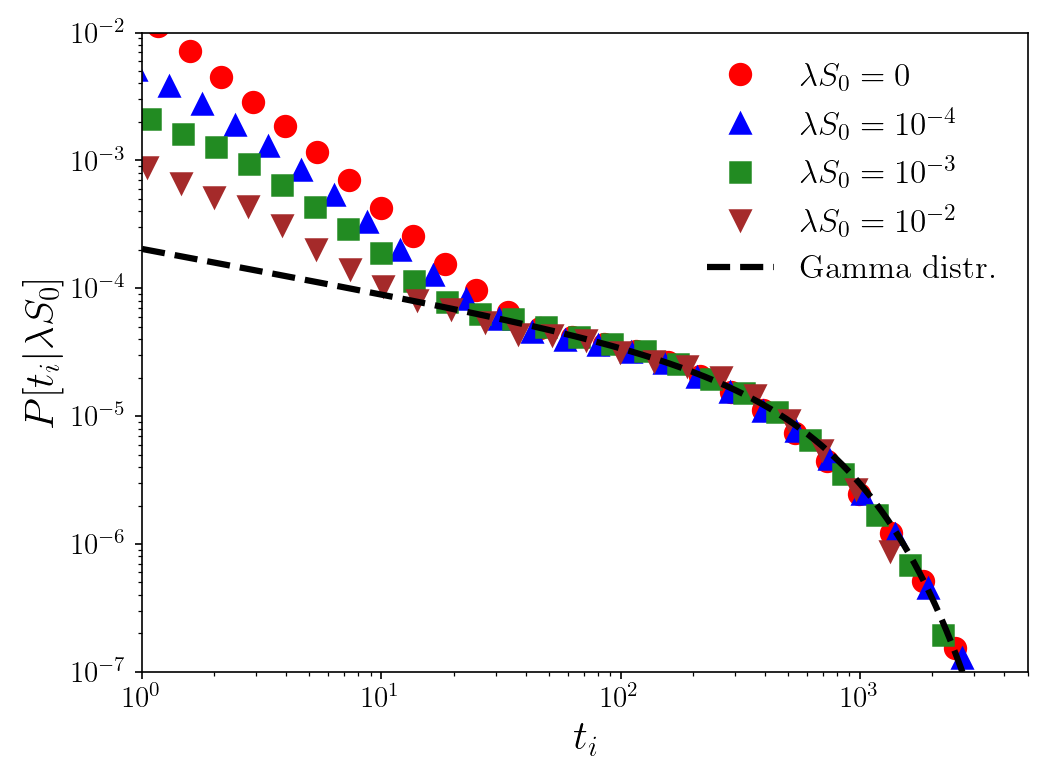}
    \caption{Probability density $P[t_i | \lambda S_0]$ for different values of $\lambda S_0$, for the space-dependent fluidity model Eqs. \eqref{eq:maxwell_model} and \eqref{eq:fluidity_eq} with $\bar \delta= 0.15$.  The black dashed line is the Gamma distribution with $\gamma = 0.35$. Numerical results obtained with same parameters as in Fig.~\ref{fig_sup:1d_ti}.}
    \label{fig_sup:1d_coarsening}
\end{figure}

For granular systems it was found that when the packing ratio of the emulsions was relatively low, or when the system has the ability to flow, there was no long tail present in the interevent time statistics \cite{Kumar_Korkolis_Benzi_Denisov_Niemeijer_Schall_Toschi_Trampert_2020} . As argued before, to model the ability to flow, or have a shear band, we can impose the boundary condition $\phi(L) = m(\Sigma)$. In Fig.~\ref{fig_sup:1d_ti} we find that for this choice of boundary condition, there is no long tail present in the interevent time statistics.
\label{section:1d_model}
We can further show that the probability distribution of the interevent time is also invariant under thresholding on the avalanche size. This is depicted in Fig.~\ref{fig_sup:1d_coarsening}, where we see that for $\bar \delta = 0.15$, the $P[t_i | \lambda S_0]$ collapses to a Gamma distribution with $\gamma=0.35$ for different $\lambda S_0$. This implies that the same analysis as done for the simplified fluidity model can be generalized when taking spatial fluctuations in the fluidity into account.

\subsection{Changing the definition of the size}\label{sec:changing_definition}
\begin{figure}[h!]
    \centering
    \includegraphics[width=0.89\linewidth]{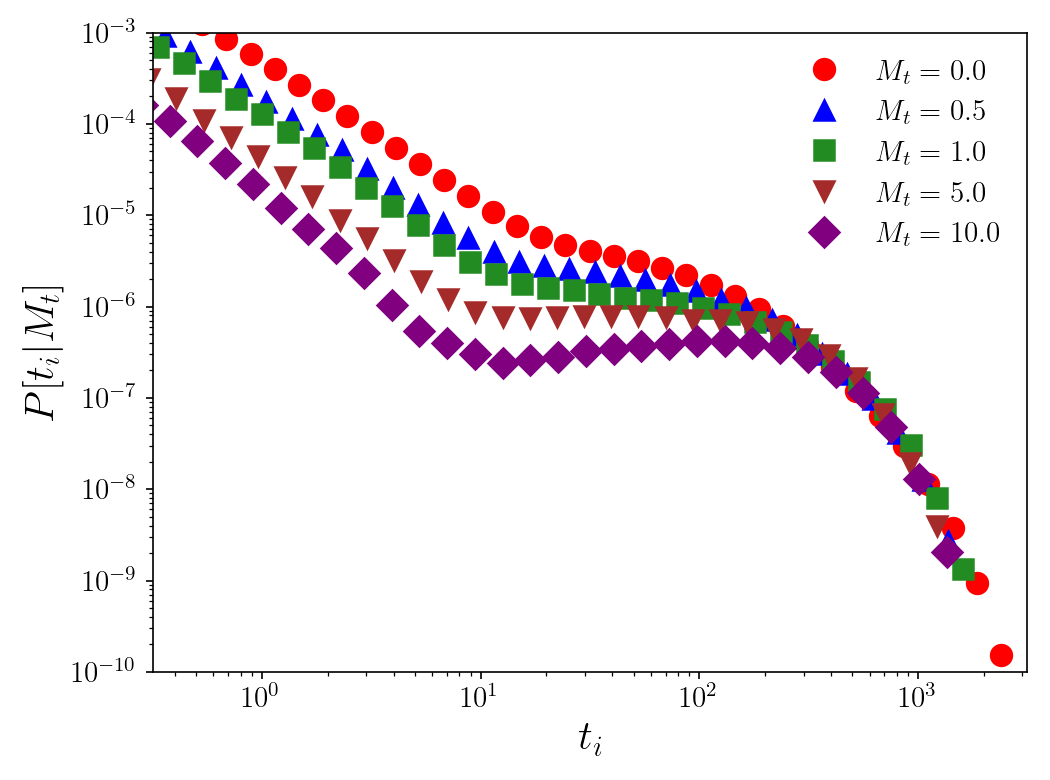}
    \caption{Probability density $P[t_i|M_{t}]$ for different values of $M_{t}$, where an avalanche is defined to occur when $\frac{d\Sigma}{dt} < -M_{t}$. Numerics done for model described by Eqs.~\eqref{eq:maxwell_model} and~\eqref{eq:0d_fluidity} with same parameters as in Fig. \ref{fig:pdf_ti_sizes}.}
    \label{fig:changing_size_definition}
\end{figure}

In our main work we have defined an avalanche to happen when $\frac{d \Sigma}{dt} < 0$. In the previous section we have applied a threshold on the size of the avalanche, showing that the probability distribution $P[t_i |\lambda S_0]$ was invariant under this transformation. Looking again at the statistics of the interevent time for the zero-dimensional model given by Eqs.~\eqref{eq:maxwell_model} and~\eqref{eq:0d_fluidity},  we now show that the probability distribution $P[t_i|M_{t}]$ is not invariant when the definition of the avalanche is changed. Instead of coarse-graining the signal, we will change the definition of the start of an avalanche to be $\frac{d \Sigma}{dt} < -M_{t}$. Thus we recover our previous results for $M_{t} = 0$. In Fig.~\ref{fig:changing_size_definition} it can be seen that the probability distribution $P[t_i|M_{t}]$ is not invariant under this thresholding in the range $\sim 10-500$ range. This shows that changing the definition from $M_{t} \rightarrow\lambda M_{t}$ is not the same kind of coarse-graining transformation as seen in the main part of the paper, since we are looking at different events~\cite{Benzi_Castaldi_Toschi_Trampert_2022}. 
\newpage

\end{document}